\documentclass[twocolumn,floatfix,showpacs,superscriptaddress,prb]{revtex4}
\usepackage{graphicx}
\usepackage{amsmath}
\usepackage{bm}
\begin{document}

\title{Josephson effect in ballistic graphene}
\author{M. Titov}
\affiliation{Department of Physics, Konstanz University, D--78457 Konstanz, Germany}
\author{C. W. J. Beenakker}
\affiliation{Instituut-Lorentz, Universiteit Leiden, P.O. Box 9506, 2300 RA Leiden, The Netherlands}
\date{May 2006}
\begin{abstract}
We solve the Dirac-Bogoliubov-De-Gennes equation in an impurity-free superconductor-normal-superconductor (SNS) junction, to determine the maximal supercurrent $I_{c}$ that can flow through an undoped strip of graphene with heavily doped superconducting electrodes. The result $I_{c}\simeq(W/L)e\Delta_{0}/\hbar$ is determined by the superconducting gap $\Delta_{0}$ and by the aspect ratio of the junction (length $L$, small relative to the width $W$ and to the superconducting coherence length). Moving away from the Dirac point of zero doping, we recover the usual ballistic result $I_{c}\simeq (W/\lambda_{F})e\Delta_{0}/\hbar$, in which the Fermi wave length $\lambda_{F}$ takes over from $L$. The product $I_{c}R_{\rm N}\simeq \Delta_{0}/e$ of critical current and normal-state resistance retains its universal value (up to a numerical prefactor) on approaching the Dirac point.
\end{abstract}
\maketitle

While the Josephson effect was originally discovered in a tunnel junction,\cite{Jos64} any weak link between two superconductors can support a dissipationless current in equilibrium.\cite{Lik79} The current $I(\phi)$ varies periodically with the phase difference $\phi$ of the pair potential in the two superconductors, reaching a maximum $I_{c}$ (the critical current) which is characteristic of the strength of the link. A measure of the coupling strength is the resistance $R_{\rm N}$ of the junction when the superconductors are in the normal state. The product $I_{c}R_{\rm N}$ increases as the separation $L$ of the two superconductors becomes smaller and smaller, until it saturates at a value of order $\Delta_{0}/e$, determined only by the excitation gap $\Delta_{0}$ in the superconductors --- but independent of the coupling strength. This phenomenology has been well established in a variety of superconductor--normal-metal--superconductor (SNS) junctions\cite{Tin04} and forms the basis of operation of the Josephson-Field-Effect-Transistor.\cite{Aka94,Mor98}

A new class of weak links has now become available for research,\cite{Nov04} in which the superconductors are coupled by a monoatomic layer of carbon (= graphene). The low-lying excitations in this material are described by a relativistic wave equation, the Dirac equation. They are massless, having a velocity $v$ that is independent of energy, and gapless, occupying conduction and valence bands that touch at discrete points (= Dirac points) in reciprocal space. \cite{PT06} Graphene thus provides a unique opportunity to explore the physics of the ``relativistic Josephson effect'' (which had remained unexplored in earlier work\cite{Kap99} on relativistic effects in high-temperature and heavy-fermion superconductors). We address this problem here in the framework of the Dirac-Bogoliubov-De-Gennes (DBdG) equation of Ref.\ \onlinecite{Bee06}.

The basic question that we seek to answer is what happens to the critical current as we approach the Dirac point of zero carrier concentration. Earlier theories\cite{Kat05,Two06,Zie06} have found that undoped graphene has a quantum-limited conductivity of order $e^{2}/h$, in the absence of any impurities or lattice defects. We find that the critical current is given, up to numerical coefficients of order unity, by
\begin{equation}
I_{c}\simeq\frac{e\Delta_{0}}{\hbar}\,\max(W/L,W/\lambda_{F}),\label{Icresult}
\end{equation}
in the short-junction regime $L\ll W,\xi$ (with $\xi=\hbar v/\Delta_{0}$ the superconducting coherence length, $W$ the width of the junction, and $\lambda_{F}$ the Fermi wave length in the normal region). At the Dirac point $\lambda_{F}\rightarrow\infty$, so the critical current reaches its minimal value of $(e\Delta_{0}/\hbar)\times W/L$. Since the normal-state resistance has its maximal value $R_{\rm N}\simeq (h/e^{2})\times L/W$ at the Dirac point, the $I_{c}R_{\rm N}$ product remains of order $\Delta_{0}/e$ as the carrier concentration is reduced to zero. 

\begin{figure}[tb]
\centerline{\includegraphics[width=0.8\linewidth]{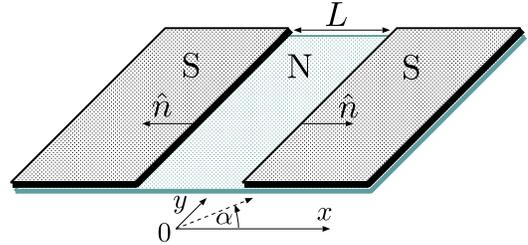}}
\caption{\label{Jojunction}
Schematic of a graphene layer, partially covered by two superconducting electrodes (S). A dissipationless supercurrent flows in equilibrium through the normal region (N), depending on the phase difference between the two superconductors. Separate gate electrodes (not shown) make it possible to vary independently the carrier concentration in the normal and superconducting regions of the graphene layer.
}
\end{figure}

The system considered is shown schematically in Fig.\ \ref{Jojunction}. A layer of graphene in the $x-y$ plane is covered by superconducting electrodes in the regions $x<-L/2$ and $x>L/2$. The normal region $|x|<L/2$ has electron and hole excitations described by the DBdG equation,\cite{Bee06}
\begin{equation}
\begin{pmatrix}
H_0-\mu & 0\\ 
0 & \mu-H_0
\end{pmatrix}
\begin{pmatrix}
\Psi_{e}\\ \Psi_{h}
\end{pmatrix}
=\varepsilon
\begin{pmatrix}
\Psi_{e}\\ \Psi_{h}
\end{pmatrix}.\label{H}
\end{equation}
Here $H_{0}=-i \hbar v(\sigma_x\partial_{x}+\sigma_{y}\partial_{y})$ is the Dirac Hamiltonian, $\varepsilon>0$ is the excitation energy, and $\mu$ is the chemical potential or Fermi energy in the normal region (measured with respect to the Dirac point, so that $\mu=0$ corresponds to undoped graphene). The electron wave functions $\Psi_{e}$ and the hole wave functions $\Psi_{h}$ have opposite spin and valley indices, which are not written explicitly. (A four-fold degeneracy factor will be added in the final results.) The Pauli matrices $\sigma_{i}$ in $H_{0}$ operate on the isospin index, which labels the two sublattices of the honeycomb lattice of carbon atoms.

Andreev reflection at a normal-superconductor (NS) interface couples $\Psi_{e}$ and $\Psi_{h}$. This coupling may be described {\em globally\/} by a scattering matrix, as was done in Ref.\ \onlinecite{Bee06} to determine the conductance of an NS junction. Here we follow a different approach, more suited to determine the energy spectrum (and therefrom the Josephson current). In this approach electrons and holes are coupled {\em locally\/} by means of a boundary condition on the wave function in the normal region. 

We consider the energy range $\varepsilon<\Delta_{0}$ below the excitation gap $\Delta_{0}$ in the superconductor, where the spectrum is discrete. At a point $\bm{r}$ on the NS interface (with unit vector $\hat{\bm{n}}$ pointing from N to S, perpendicular to the interface), the boundary condition takes the form
\begin{eqnarray}
&&\Psi_{h}(\bm{r})=M\Psi_{e}(\bm{r}),\label{A}\\
&&M=\frac{1}{\Delta}\bigl(\varepsilon -i\sqrt{|\Delta|^{2}-\varepsilon^{2}}\,\hat{\bm{n}}\cdot{\bm\sigma}\bigr)=e^{-i\Phi-i\beta\,\hat{\bm{n}}\cdot{\bm\sigma}}.\label{Mdef}
\end{eqnarray}
Here $\Delta=\Delta_{0}e^{i\Phi}$ is the complex pair potential in S, ${\bm \sigma}=(\sigma_{x},\sigma_{y})$ is the vector of Pauli matrices, and $\beta=\arccos(\varepsilon/\Delta_{0})\in(0,\pi/2)$.

The relation (\ref{A}) follows from the DBdG equation,\cite{Bee06,note1} under three assumptions characterizing an ``ideal'' NS interface: I) The Fermi wave length $\lambda'_{F}$ in S is sufficiently small that $\lambda'_{F}\ll\xi,\lambda_{F}$, where $\lambda_{F}=\hbar v/\mu$ is the Fermi wave length in N and $\xi=\hbar v/\Delta_{0}$ is the superconducting coherence length;  II) The interface is smooth and impurity free on the scale of $\xi$; III) There is no lattice mismatch at the NS interface, so the honeycomb lattice of graphene is unperturbed at the boundary. The absence of lattice mismatch might be satisfied by depositing the superconductor on top of a heavily doped region of graphene. As in the case of a semiconductor two-dimensional electron gas,\cite{Vol95,Fag05} we expect that such an extended superconducting contact can be effectively described by a pair potential $\Delta$ in the $x-y$ plane (even though graphene by itself is not superconducting).

The particle current density out of the normal region, given by
\begin{equation}
\bm{j}_{\rm particle}=v\Psi_{e}^{\ast}\hat{\bm{n}}\cdot\bm{\sigma}\Psi_{e}-v\Psi_{h}^{\ast}\hat{\bm{n}}\cdot\bm{\sigma}\Psi_{h},\label{jpartdef}
\end{equation}
should vanish for $\varepsilon<\Delta_{0}$, because subgap excitations decay over a length $\xi$ in S. (The possibility of a subgap excitation entering the superconductor at one point along the boundary and exiting at another point within a distance $\xi$ is excluded by assumption II.) By substituting the boundary condition (\ref{A}) one indeed finds that $\bm{j}_{\rm particle}=0$, since $M$ is a unitary matrix which commutes with $\hat{\bm{n}}\cdot{\bm\sigma}$.

In the SNS junction the normal region has two interfaces with the superconductor, one at $x=-L/2$ (with superconducting phase $\Phi=\phi/2$ and outward normal $\hat{\bm{n}}=-\hat{\bm{x}}$) and another at $x=L/2$ (with $\Phi=-\phi/2$ and $\hat{\bm{n}}=\hat{\bm{x}}$). The boundary condition (\ref{A}) at the points $\bm{r}_{\pm}=(\pm L/2,y)$ thus takes the form
\begin{eqnarray}
&&\Psi_{h}(\bm{r}_{-})=U(\varepsilon)\Psi_{e}(\bm{r}_{-}),\;\;\Psi_{h}(\bm{r}_{+})=U^{-1}(\varepsilon)\Psi_{e}(\bm{r}_{+}),\label{A_SNS}\\
&&U(\varepsilon)=e^{-i\phi/2+i\beta\sigma_{x}},\;\;\beta=\arccos(\varepsilon/\Delta_{0}).\label{Udef}
\end{eqnarray}

Since the wave vector $k_{y}$ parallel to the NS interface is conserved upon Andreev reflection, we may solve the problem for a given $k_{y}\equiv q$. The transfer matrix ${\cal M}(\varepsilon,q)$ relates the states at the two ends of the normal region: 
\begin{equation}
\Psi_{e}(\bm{r}_{+})={\cal M}(\varepsilon,q)\Psi_{e}(\bm{r}_{-}),\;\;
\Psi_{h}(\bm{r}_{+})={\cal M}(-\varepsilon,q)\Psi_{h}(\bm{r}_{-}).
\end{equation}
(For ease of notation, the $q$-dependence will not be written explicitly in what follows.) The condition for a bound state (= Andreev level) in the SNS junction is that the transfer matrix for the round-trip from $\bm{r}_{-}$ to $\bm{r}_{+}$ and back to $\bm{r}_{-}$ has an eigenvalue equal to unity. This condition can be written in the form of a determinant,
\begin{equation}
{\rm Det}\,\bigl[1-{\cal M}^{-1}(\varepsilon)U(\varepsilon){\cal M}(-\varepsilon)U(\varepsilon)\bigr]=0,\label{Detequation}
\end{equation}
which we have to solve for $\varepsilon$ as a function of $q$ and $\phi$. 

The electron transfer matrix ${\cal M}(\varepsilon)$ is readily obtained from the Dirac equation,
\begin{eqnarray}
&&{\cal M}=\Lambda e^{ikL\sigma_{z}}\Lambda,\label{Mresult}\\
&&\Lambda=\Lambda^{-1}=(2\cos\alpha)^{-1/2}
\begin{pmatrix}
e^{-i\alpha/2}&e^{i\alpha/2}\\
e^{i\alpha/2}&-e^{-i\alpha/2}
\end{pmatrix},\label{Lambdadef}\\
&&\alpha(\varepsilon)=\arcsin\left(\frac{\hbar vq}{\varepsilon+\mu}\right),\label{alphadef}\\
&&k(\varepsilon)=(\hbar v)^{-1}(\varepsilon+\mu)\cos\alpha(\varepsilon).\label{kdef}
\end{eqnarray}
The angle $\alpha$ is the angle of incidence of the electron, and $k$ is its longitudinal wave vector.

Evaluation of the determinant (\ref{Detequation}) leads after some algebra to the quantization condition
\begin{eqnarray}
\cos\phi&=&\left(\cos\theta_{+}\cos\theta_{-}+\frac{\sin\theta_{+}\sin\theta_{-}}{\cos\alpha_{+}\cos\alpha_{-}}\right)\cos 2\beta\nonumber\\
&&\mbox{}+\left(\frac{\sin\theta_{+}\cos\theta_{-}}{\cos\alpha_{+}}-\frac{\cos\theta_{+}\sin\theta_{-}}{\cos\alpha_{-}}\right)\sin 2\beta\nonumber\\
&&\mbox{}-\sin\theta_{+}\sin\theta_{-}\tan\alpha_{+}\tan\alpha_{-},\label{qc1}
\end{eqnarray}
where we abbreviated $\alpha_{\pm}=\alpha(\pm\varepsilon)$, $\theta_{\pm}=k(\pm\varepsilon)L$. 

We introduce a finite width $W$ to quantize the transverse wave vectors, $q\rightarrow q_{n}$, $n=0,1,2,\ldots$, and denote by $\rho_{n}(\varepsilon,\phi)$ the density of states in mode $n$. The Josephson current at zero temperature is then given by
\begin{equation}
I(\phi)=-\frac{4e}{\hbar}\frac{d}{d\phi}\int_{0}^{\infty}d\varepsilon\,\sum_{n=0}^{\infty}\rho_{n}(\varepsilon,\phi)\,\varepsilon,
\label{Iphi}
\end{equation}
where the factor of $4$ accounts for the two-fold spin and valley degeneracies. To be definite we take ``infinite mass'' boundary conditions at $y=0,W$, for which\cite{Two06} $q_{n}=(n+1/2)\pi/W$. (For $W\gg L$ the choice of boundary conditions becomes irrelevant.) At the Fermi level, the lowest $N(\mu)=\mu W/\pi\hbar v$ modes are propagating (real $k$), while the higher modes are evanescent (imaginary $k$).

We analyze the Josephson effect in the experimentally most relevant {\em short-junction\/} regime that the length $L$ of the normal region is small relative to the superconducting coherence length $\xi$. In terms of energy scales, this condition requires $\Delta_{0}\ll\hbar v/L$. To leading order in the small parameter $\Delta_{0}L/\hbar v$ we may substitute $\alpha_{\pm}\rightarrow\alpha(0)$, $\theta_{\pm}\rightarrow k(0)L$ in the quantization condition (\ref{qc1}). The solution is a single bound state per mode,
\begin{eqnarray}
&&\varepsilon_{n}(\phi)=\Delta_{0}\sqrt{1-\tau_{n}\sin^{2}(\phi/2)},\label{epsshort}\\
&&\tau_{n}=\frac{k_{n}^{2}}{k_{n}^{2}\cos^{2}(k_{n}L)+\mu^{2}\sin^{2}(k_{n}L)},\label{taudef}
\end{eqnarray}
with $k_{n}=[(\mu/\hbar v)^{2}-q_{n}^{2}]^{1/2}$. This expression for the Andreev levels in terms of a normal-state transmission probability $\tau_{n}$ has the usual form for a short SNS junction.\cite{Bee91} Comparison with Ref.\ \onlinecite{Two06} shows that $\tau_{n}$ is indeed the transmission probability for a ballistic strip of graphene between two heavily doped electrodes in the normal state ($\Delta_{0}=0$, $\lambda'_{F}\ll\lambda_{F}$).  The normal-state resistance $R_{\rm N}$ is thus given by
\begin{equation}
R_{\rm N}^{-1}=\frac{4e^{2}}{h}\sum_{n=0}^{\infty}\tau_{n}.\label{RNdef}
\end{equation}

\begin{figure}[tb]
\centerline{\includegraphics[width=0.9\linewidth]{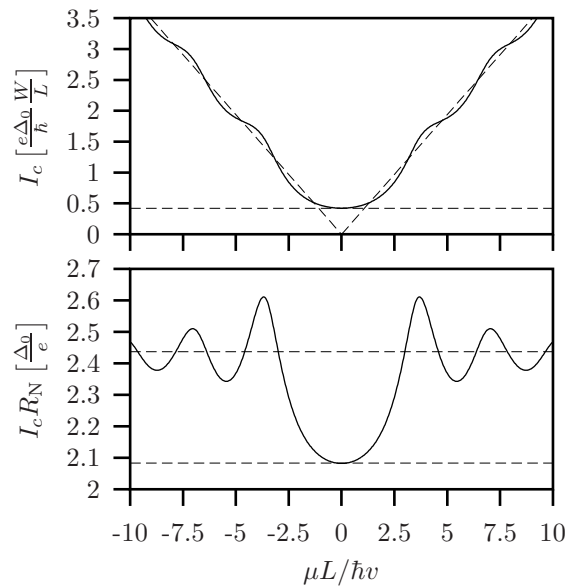}}
\caption{\label{Icplot}
Critical current $I_{c}$ and $I_{c}R_{\rm N}$ product of a ballistic Josephson junction (length $L$ short compared to the width $W$ and superconducting coherence length $\xi$), as a function of the Fermi energy $\mu$ in the normal region. The asymptotes (\ref{IcDirac}) and (\ref{Icmu}) are indicated by dashed lines.
}
\end{figure}

Substitution of $\rho_{n}(\varepsilon,\phi)=\delta[\varepsilon-\varepsilon_{n}(\phi)]$ into Eq.\ (\ref{Iphi}) gives the supercurrent due to the discrete spectrum,
\begin{equation}
I(\phi)=\frac{e\Delta_{0}}{\hbar}\sum_{n=0}^{\infty}\frac{\tau_{n}\sin\phi}{[1-\tau_{n}\sin^{2}(\phi/2)]^{1/2}}.\label{Iphishorthigh}
\end{equation}
Contributions to the supercurrent from the continuous spectrum are smaller by a factor $L/\xi$ and may be neglected in the short-junction regime.\cite{Bee92} For $L\ll W$ the summation over $n$ may be replaced by an integration. The resulting critical current $I_{c}$ and the $I_{c}R_{\rm N}$ product are plotted as a function of $\mu$ in Fig.\ \ref{Icplot}.

The limiting behavior at the Dirac point ($\mu\ll\hbar v/L$) for a short and wide normal region ($L\ll W,\xi$) is
\begin{eqnarray}
&&I(\phi)=\frac{e\Delta_{0}}{\hbar}\,\frac{2W}{\pi L}\cos(\phi/2)\,{\rm artanh}\,[\sin(\phi/2)],\label{IphiDirac}\\
&&I_{c}=1.33\,\frac{e\Delta_{0}}{\hbar}\,\frac{W}{\pi L},\;\;I_{c}R_{\rm N}=2.08\,\Delta_{0}/e.\label{IcDirac}
\end{eqnarray}
These results for {\em ballistic\/} graphene at the Dirac point are formally identical to those of a {\em disordered\/} normal metal (Fermi wave vector $k_{F}$, mean free path $l$),\cite{Bee91,Kul75} upon substitution $k_{F}l\rightarrow 1$. This correspondence is consistent with the finding of Ref.\ \onlinecite{Two06}, that ballistic Dirac fermions have the same shot noise as diffusive nonrelativistic electrons.

In the opposite regime $\mu\gg\hbar v/L$ we have instead (still for $L\ll W,\xi$) the result
\begin{equation}
I_{c}=1.22\,\frac{e\Delta_{0}}{\hbar}\,\frac{\mu W}{\pi\hbar v},\;\;I_{c}R_{\rm N}=2.44\,\Delta_{0}/e.\label{Icmu}
\end{equation}
(We do not have a simple analytic expression for the $\phi$-dependence in this regime.) The critical current (\ref{Icmu}) is about half the ideal ballistic value\cite{Bee91,Kul77} $I_{c}=2Ne\Delta_{0}/\hbar$, with $N=\mu W/\pi\hbar v$ the number of propagating modes (per spin and valley). This reduction is due to the mismatch in Fermi wave length at the NS interfaces. Eqs.\  (\ref{IcDirac}) and (\ref{Icmu}) together contain the scaling behavior (\ref{Icresult}) announced in the introduction. 

In conclusion, we have shown that a Josephson junction in graphene can carry a nonzero supercurrent even if the Fermi level is tuned to the point of zero carrier concentration. At this Dirac point, the current-phase relationship has the same form as in a disordered normal metal --- but without any impurity scattering. Instead of being independent of the length $L$ of the junction, as expected for a short ballistic Josephson junction, the critical current $I_{c}$ at the Dirac point has the diffusion-like scaling $\propto 1/L$. Since the normal-state resistance $R_{\rm N}\propto L$, the $I_{c}R_{\rm N}$ product remains fixed at the superconducting gap (up to a numerical prefactor) as the Fermi level passes through the Dirac point. This unusual ``quasi-diffusive'' scaling of the Josephson effect in undoped graphene should be observable in submicron scale junctions. 

This research was supported by the Dutch Science Foundation NWO/FOM. We have benefitted from discussions on the experimental implications of this work with A. Morpurgo, B. Trauzettel, L. M. K. Vandersypen, and other members of the Delft/Leiden focus group on Solid State Quantum Information Processing.


\begin{thebibliography}{99}
\bibitem{Jos64} B. D. Josephson, Rev.\ Mod.\ Phys.\ {\bf 36}, 216 (1964).
\bibitem{Lik79} K. K. Likharev, Rev.\ Mod.\ Phys.\ {\bf 51}, 101 (1979).
\bibitem{Tin04} M. Tinkham, {\em Introduction to Superconductivity\/} (Dover, New York, 2004).
\bibitem{Aka94} T. Akazaki, J. Nitta, H. Takayanagi, T. Enoki, and K. Arai, Appl.\ Phys.\ Lett.\ {\bf 65}, 1263 (1994).
\bibitem{Mor98} A. F. Morpurgo, T. Klapwijk, and B. J. van Wees, Appl.\ Phys.\ Lett.\ {\bf 72}, 966 (1998).
\bibitem{Nov04} K. S. Novoselov, A. K. Geim, S. V. Morozov, D. Jiang, Y. Zhang, S. V. Dubonos, I. V. Grigorieva, and A. A. Firsov, Science {\bf 306}, 666 (2004).
\bibitem{PT06} M. Wilson, Physics Today, January 2006, p.\ 21.
\bibitem{Kap99} K. Capelle and E. K. U. Gross, Phys.\ Rev.\ B {\bf 59}, 7140 (1999); {\bf 59}, 7155 (1999).
\bibitem{Bee06} C. W. J. Beenakker, cond-mat/0604594.
\bibitem{Kat05} M. I. Katsnelson, cond-mat/0512337.
\bibitem{Two06} J. Tworzyd{\l}o, B. Trauzettel, M. Titov, A. Rycerz, and C. W. J. Beenakker, cond-mat/0603315.
\bibitem{Zie06} K. Ziegler, cond-mat/0604537.
\bibitem{note1} The boundary condition (\ref{A}) follows from the general solution of the DBdG equation near a planar NS interface of Ref.\ \onlinecite{Bee06} (Appendix A). For $\lambda'_{F}\ll\xi,\lambda_{F}$ the wave functions satisfy Eq.\ (\ref{A}) at all points in S. Continuity of the wave function at the NS interface then implies that this relation holds as a boundary condition for the wave functions in N. The same Eq.\ (\ref{A}) also holds for $\varepsilon>\Delta_{0}$, provided that there are no excitations propagating from S into N. The angle $\beta$ is then imaginary, $\beta=-i\,{\rm arcosh}\,(\varepsilon/\Delta_{0})$.
\bibitem{Vol95} A. F. Volkov, P. H. C. Magnee, B. J. van Wees, and T. M. Klapwijk, Physica C {\bf 242}, 261 (1995).
\bibitem{Fag05} G. Fagas, G. Tkachov, A. Pfund, and K. Richter, Phys.\ Rev.\ B {\bf 71}, 224510 (2005).
\bibitem{Bee91} C. W. J. Beenakker, Phys.\ Rev.\ Lett.\ {\bf 67}, 3836 (1991); {\bf 68}, 1442(E) (1992); extended version in: {\em Transport Phenomena in Mesoscopic Systems}, edited by H. Fukuyama and T. Ando (Springer, Berlin, 1992; online at cond-mat/0406127).
\bibitem{Bee92} C. W. J. Beenakker and H. van Houten, Phys.\ Rev.\ Lett.\ {\bf 66}, 3056 (1991); extended version in: {\em Nanostructures and Mesoscopic Systems}, edited by W. P. Kirk and M. A. Reed (Academic, New York, 1992; online at cond-mat/0512610).
\bibitem{Kul75} I. O. Kulik and A. N. Omel'yanchuk, JETP Lett.\ {\bf 21}, 96 (1975).
\bibitem{Kul77} I. O. Kulik and A. N. Omel'yanchuk, Sov.\ J. Low Temp.\ Phys.\ {\bf 3}, 459 (1977); {\bf 4}, 142 (1978).
\end{thebibliography}
\end{document}